\documentclass[twocolumn,english,aps, prb, showpacs]{revtex4}
\usepackage[T1]{fontenc}
\usepackage[latin9]{inputenc}
\usepackage{float}
\usepackage{amsbsy}
\usepackage{amstext}
\usepackage{graphicx}
\usepackage{esint}
\usepackage{babel}

\begin{document}

\title{Magnetic impurity in a $U(1)$-Spin Liquid with a Spinon Fermi-Surface}

\author{P. Ribeiro}
\email{ribeiro@cfif.ist.utl.pt}
\affiliation{CFIF, Instituto Superior Técnico, Universidade Técnica de Lisboa,
Av. Rovisco Pais, 1049-001 Lisboa, Portugal}

\author{P. A. Lee}
\affiliation{Department of Physics, Massachusetts Institute of Technology, Cambridge,
Massachusetts 02139}

\begin{abstract}
We address the problem of a magnetic impurity in a two dimensional
$U(1)$ spin liquid where the spinons have gap-less excitations near
the Fermi-surface and are coupled to an emergent gap-less gauge field.
Using a large N expansion we analyze the strong coupling behavior
and obtain the Kondo temperature which was found to be the same as
for a Fermi-liquid. In this approximation we also study the specific
heat and the magnetic susceptibility of the impurity. These quantities
present no deviations from the Fermi-liquid ones, consistent with
the notion that the magnetic impurity is only sensitive to the local
density of fermionic states.
\end{abstract}

\maketitle

\global\long\def\ket#1{\left| #1\right\rangle }
\global\long\def\bra#1{\left\langle #1 \right|}
\global\long\def\kket#1{\left\Vert #1\right\rangle }
\global\long\def\bbra#1{\left\langle #1\right\Vert }
\global\long\def\braket#1#2{\left\langle #1\right. \left| #2 \right\rangle }
\global\long\def\bbrakket#1#2{\left\langle #1\right. \left\Vert #2\right\rangle }
\global\long\def\av#1{\left\langle #1 \right\rangle }
\global\long\def\tr{\text{Tr}}
\global\long\def\im{\text{Im}}
\global\long\def\re{\text{Re}}
\global\long\def\sign{\text{sgn}}
\global\long\def\abs#1{\left|#1\right|}

\section{introduction}

A large number of theoretical proposals for the low-energy description
of spin-liquid phases consider fractionalized fermionic degrees of
freedom, the spinons, carrying spin $1/2$ but no electric charge,
coupled to an emergent $U(1)$ gauge field, a gap-less photon-like
mode. The spinons are gap-less having either nodal points \cite{Wen_2002}
or a Fermi-surface. The former case arises naturally in the slave
particle approach to the $t-J$ model \cite{Lee_1992} but also in
other physical contexts such as the half-filled Landau level \cite{Halperin_1993}
and the description of metals at a Pomeranchuk instability \cite{Oganesyan_2001}.
It presents non-Fermi liquid behavior due to the strong interactions
between the spinons and the gauge field that lead to a spectrum with
no well defined quasi-particles. This phase has a number of remarkable
thermodynamical and transport properties \cite{Nave_2007_b}, for
example at low temperatures soft gauge modes contribute to the specific
heat with a term proportional to $T^{2/3}\ $ \cite{Motrunich_2005}.\\
\\
Magnetic impurities embedded in a parent material provide an experimental
probe to the bulk properties and can help to discriminate between
possible candidate phases. Moreover in order to be observed experimentally
the system itself should be stable to a dilute density of such impurities.
In a Fermi-liquid an antiferromagnetically coupled spin impurity leads
to the well known Kondo effect \cite{Hewson_1996} characterized by
a cross over from the low temperature strong coupled regime, where
the magnetic moment of the impurity is completely screened by the
bulk quasi-particles, to the high temperature regime, where the impurity
susceptibility follows a Curie-Weiss law. This cross-over occurs near
the Kondo temperature which is an example of a dynamically generated
energy scale. Since the understanding of the Kondo effect the study
of impurities in different bulk phases has attracted much attention
\cite{Cassanello_1996,Cassanello_1997,Florens_2006,Kolezhuk_2006,Kim_2008},
in particular for bosonic \cite{Florens_2006} and algebraic spin
liquids \cite{Kolezhuk_2006,Kim_2008}. \\
\\
The purpose of the present work is to study the behavior of a magnetic
impurity embedded in a $U(1)$ spin liquid with a Fermi-surface. Being
a charge insulator this system still presents a Kondo like behavior
since the spin degrees of freedom are free to screen the magnetic
impurity at low energies. The article is organized as follows: in
sec.\ref{sec:methods} we describe the model and give some details
of the $1/N$ expansion (sec. \ref{sub:Large-N-expansion} and \ref{sub:Fluctuations}),
the specific heat and the local spin susceptibilities are respectively
computed in sec. \ref{sub:Specific-Heat-Capacity} and \ref{sub:Spin-Susceptibility}.
Finally in sec.\ref{sec:Discussion} we conclude discussing the implications
of our results.

\section{methods\label{sec:methods}}

Starting from the $t-J$ model in 2-D, the action describing the spin-liquid
phase with a spinon Fermi surface coupled to a compact $U(1)$ gauge
field can be obtained within the slave-boson formalism \cite{Lee_1992}
or using a slave-rotor representation \cite{Lee_2005}, when fluctuations
around the mean-field solution are considered. We assume that due
to the presence of a large number of gap-less fermions the system
is deconfined i.e. one can consider a non-compact $U(1)$ gauge theory
\cite{Hermele_2004}. The partition function writes as a path integral
over the spinon grassmanian fields $f_{\sigma=\pm1}$ and the bosonic
gauge fields $a_{\mu}=\left(a_{0},\boldsymbol{a}\right)$ with action
\begin{eqnarray}
\mathcal{S}_{SL} & = & \int d^{3}x\sum_{\sigma=\pm}\left\{ f_{\sigma}^{\dagger}\left(\partial_{\tau}+ia_{0}\right)f_{\sigma}\right.\label{eq:SL}\\
 &  & \left.+\frac{1}{2m}\left[\left(\boldsymbol{\partial}-i\boldsymbol{a}\right)f_{\sigma}^{\dagger}\right]\mathbf{.}\left[\left(\boldsymbol{\partial}+i\boldsymbol{a}\right)f_{\sigma}\right]-\frac{1}{b}ia_{0}\right\} ,\nonumber \end{eqnarray}
where $b$ is the microscopic lattice volume and $m$ the spinon mass.
The integration over the temporal component of the gauge field $a_{0}$
acts as an on-site chemical potential for the spinons enforcing $b\sum_{\alpha}f_{\alpha}^{\dagger}\, f_{\alpha}=1$.
We use the notation$\int d^{3}x=\int_{0}^{\beta}d\tau\int d^{2}x$.
\\
\\
At $\mathbf{x}=\mathbf{0}$ the interaction with the magnetic impurity
is given by $\mathcal{S}_{K}=\mathcal{S}_{Berry}+J_{K}b\int d\tau\ \mathbf{S}_{f}(\mathbf{0})\mathbf{.}\mathbf{S}$,
where $\mathcal{S}_{Berry}$ is the action of the free impurity spin,
$J_{K}$ is the Kondo coupling and $\mathbf{S}_{f}\left(\mathbf{0}\right)=f_{\alpha}^{\dagger}\left(\tau,\mathbf{0}\right)\boldsymbol{\sigma}_{\alpha,\beta}\, f_{\alpha}\left(\tau,\mathbf{0}\right)$.
Using a fermionic representation for the impurity spin $\mathbf{S}=c_{\alpha}^{\dagger}\boldsymbol{\sigma}_{\alpha,\beta}\, c_{\alpha}$,
this term writes explicitly $\mathcal{S}_{K}=\int d\tau\left\{ \sum_{\sigma}c_{\sigma}^{\dagger}\left(\partial_{\tau}+i\lambda\right)c_{\sigma}-i\lambda-J_{K}b\mathbf{S}_{f,0}\mathbf{.}\mathbf{S}\right\} ,$
where $\lambda$ is an integration parameter inserted in order to
enforce the constrain $\sum_{\alpha}c_{\alpha}^{\dagger}\, c_{\alpha}=1$.

\subsection{Large N expansion\label{sub:Large-N-expansion}}

Perturbative expansions for the Kondo problem are plagued with infrared
logarithmic divergences signaling the fact that, for low energy, the
system flows to a strong coupled fixed point where the impurity forms
a singlet with the bulk electrons. Even if resummation of the divergent
terms is possible this method is not well suited to describe the low
temperature phase. Alternatively the large $N$ expansion reproduces
the essential features of the Kondo effect in the strong coupling
regime. However for temperatures of the order of the Kondo temperature
$T_{K}$, where a cross over to the asymptotic free regime is expected,
this technique becomes unreliable due to the violation of the occupancy
constrain and instead predicts a continuous phase transition\cite{Bickers_1987}.
Therefore our results are restricted to the low energy regime. For
the $U(1)$ spin liquid the large $N$ expansion corresponds to the
random-phase-approximation (RPA) used to obtain most of the physical
predictions for this phase\cite{Nave_2007_b}. Recently the validity
of this method applied to this specific problem was questioned \cite{Lee_2009}
since all planar diagrams where shown to contribute to leading order.
A possible resolution was proposed in \cite{Mross_2010} using a double
expansion to control higher loop contributions and essentially recovering
the RPA result.\\
\\
In order to perform a saddle-point expansion we generalize the
above action to $su(N)$ following the standard procedure\cite{Read_1983,Cassanello_1996,Bickers_1987}:
the Pauli matrices $\boldsymbol{\sigma}=\{\sigma_{1},...,\sigma_{3}\}$
are replaced by the generators of $su(N)$ $\boldsymbol{\tau}=\{...,\tau^{a},...\}$
with the index $a=1,..,N^{2}-1$ and the coupling constant is rescaled
$J_{K}\to\frac{J_{K}}{N}$. The representation of the impurity spin
is taken to be conjugate to the spinons one. Using the Fierz-like
identity\cite{Cassanello_1996} the Kondo term writes\begin{eqnarray}
\mathcal{S}_{K} & = & \int d\tau\left\{ \sum_{\sigma}c_{\sigma}^{\dagger}\left(\partial_{\tau}+i\lambda\right)c_{\sigma}-i\lambda Q_{f}\right.\nonumber \\
 &  & \left.+\frac{J_{K}}{N}b\left(f_{\alpha}^{\dagger}\left(0\right)c_{\alpha}c_{\beta}^{\dagger}f_{\beta}\left(0\right)\right)+J_{K}\right\} ,\end{eqnarray}
and the last term of Eq. (\ref{eq:SL}) is now multiplied by $Q_{f}$
defined such that $b\sum_{\sigma}f_{\sigma}^{\dagger}\left(\mathbf{x}\right)f_{\sigma}\left(\mathbf{x}\right)=\sum_{\sigma}c_{\sigma}^{\dagger}c_{\sigma}=Q_{f}$
. \\
\\
Following \cite{Read_1983}, the interaction term is decoupled
inserting a bosonic Hubbard-Stratonovich field $\chi=\kappa e^{i\phi}$.
The integration over $\phi$ can be absorbed by a shift in $\lambda$
leaving a single real dynamical variable $\kappa$. The integration
over the fermionic degrees of freedom can then be performed and the
partition function writes $Z=\int DaD\lambda D\kappa e^{-N\, s_{b}}$
where \begin{eqnarray}
s_{b} & = & -\frac{1}{N}\tr\ln\left[-G^{-1}\right]-\frac{1}{N}\tr\ln\left[-F^{-1}\right]+\nonumber \\
 &  & +\int dx_{0}\left\{ \frac{1}{J_{K}}\kappa^{2}-i\lambda Q_{f}\right\} -\int d^{3}x\,\frac{Q_{f}}{Nb}ia_{0},\label{eq:bosonic_action}\end{eqnarray}
is the action for the bosonic fields $a_{\mu}$, $\lambda$ and $\kappa$
only and $F^{-1}$ and $G^{-1}$ are the inverse of the full interacting
propagators of the impurity and spinon fermions. We proceed performing
a saddle-point expansion in the large $N$ limit imposing the a static
ansatz\begin{eqnarray}
\kappa(\tau) & = & \kappa_{0}\neq0,\\
\lambda(\tau) & = & -i\varepsilon_{c},\\
a_{\mu}(x) & = & \delta_{0,\mu}\ i\mu.\end{eqnarray}
At $T=0$, the variations of the action in order to $a_{1,2}$ are
trivially zero and the ones for $a_{0}$, $\lambda$ and $\kappa$
give, respectively,\begin{eqnarray}
\frac{1}{\beta}\tr\left[G_{0}\right] & = & \frac{V}{b}\frac{Q_{f}}{N},\\
\frac{1}{\pi}\tan^{-1}\left(\frac{\Delta}{\varepsilon_{c}}\right) & = & \frac{Q_{f}}{N},\label{eq:friedel}\\
n(0)b\ln\left(\frac{\Lambda}{\sqrt{\varepsilon_{c}^{2}+\Delta^{2}}}\right) & = & \frac{1}{J_{K}}.\label{eq:kondo_T}\end{eqnarray}
The first equation fixes the chemical potential $\mu$, where $G_{0}\left(i\omega_{n},\mathbf{k}\right)=\left(i\omega_{n}-\varepsilon_{\mathbf{k}}\right)^{-1}$
is the bare propagator of the spinons with single-particle energies
$\varepsilon_{k}=\frac{1}{2m}k^{2}-\mu$. $n(0)=\frac{m}{2\pi}$ is
the spinon density of states at the Fermi level and $\Lambda$ is
a high-energy cutoff for the dispersion relation. The last two equations
were obtained by Read and Newns for the Coqblin-Schriffer Hamiltonian
\cite{Read_1983}. In the limit where $\Lambda$ is much smaller than
the Fermi energy but much larger than the other energy scales the
propagator of the impurity fermions is given by $F_{0}\left(i\omega_{n}\right)=\frac{1}{i\omega_{n}-\varepsilon_{c}+i\Delta\sign\omega_{n}}$
(where $\Delta=\pi n(0)\kappa_{0}^{2}b$) corresponding to a Lorentzian
density of states $\rho(\nu)=\frac{1}{\pi}\frac{\Delta}{\left(\nu-\varepsilon_{c}\right)^{2}+\Delta^{2}}$
(see Fig. \ref{fig:Propaga}-(a)). The saddle-point values $\varepsilon_{c}$
and $\Delta$ are thus the resonance position and the hybridization
width respectively. Identifying the phase shift of an bulk spinon
scattered by the impurity $\delta_{f}\left(\omega\right)=\tan^{-1}\left(\frac{\Delta}{\varepsilon_{c}-\omega}\right)$,
Eq. (\ref{eq:friedel}) is a particular example of the Friedel sum
rule. Finally Eq. (\ref{eq:kondo_T}) defines the Kondo energy scale
$k_{B}T_{K}=\sqrt{\varepsilon_{c}^{2}+\Delta^{2}}=\Lambda e^{-\frac{1}{b\, n(0)J_{K}}}$.
At zero order in $1/N$ there is no influence of the gauge field in
the dynamics of the impurity. \\
\\
A comment about the procedure is in order at this point. One could
imagine starting with the bulk theory fixed point obtained in ref.\cite{Mross_2010},
this would correspond to first renormalize the bulk system propagators
and then introduce the impurity. However since $1/N$ is the small
parameter of our expansion entering in both the spinon and the impurity
Hamiltonians it is natural to start with the bare bulk action. The
equivalence of both results can be checked replacing the bare spinon
propagator by the integrating one.

\subsection{Fluctuations \label{sub:Fluctuations}}

\begin{figure}
\includegraphics[width=1\columnwidth]{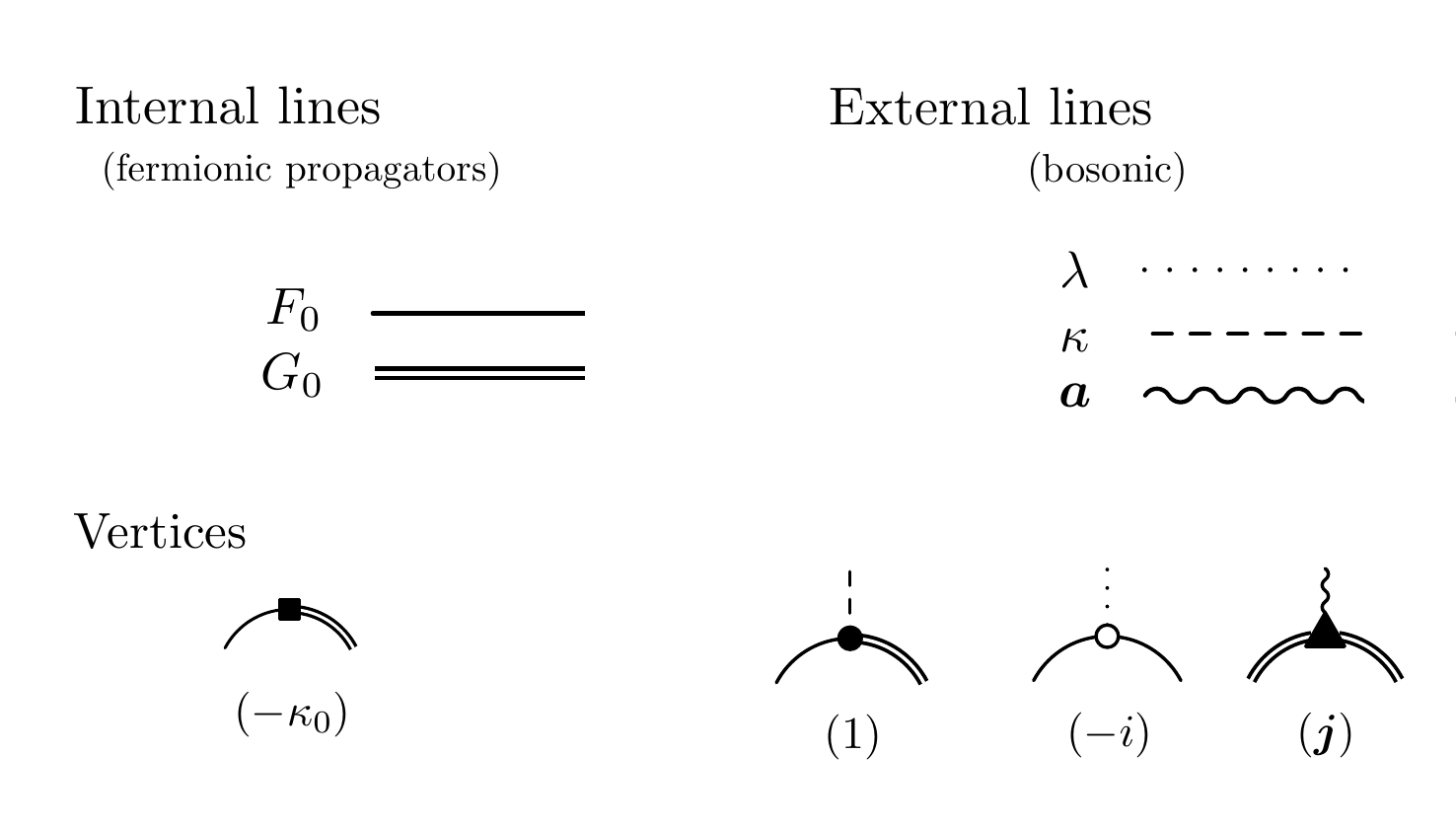}\caption{\label{fig:Vertices}Propagators, vertices and external lines used
to obtain the bubble-like diagrams. }

\end{figure}

\begin{figure}
\includegraphics[width=1\columnwidth]{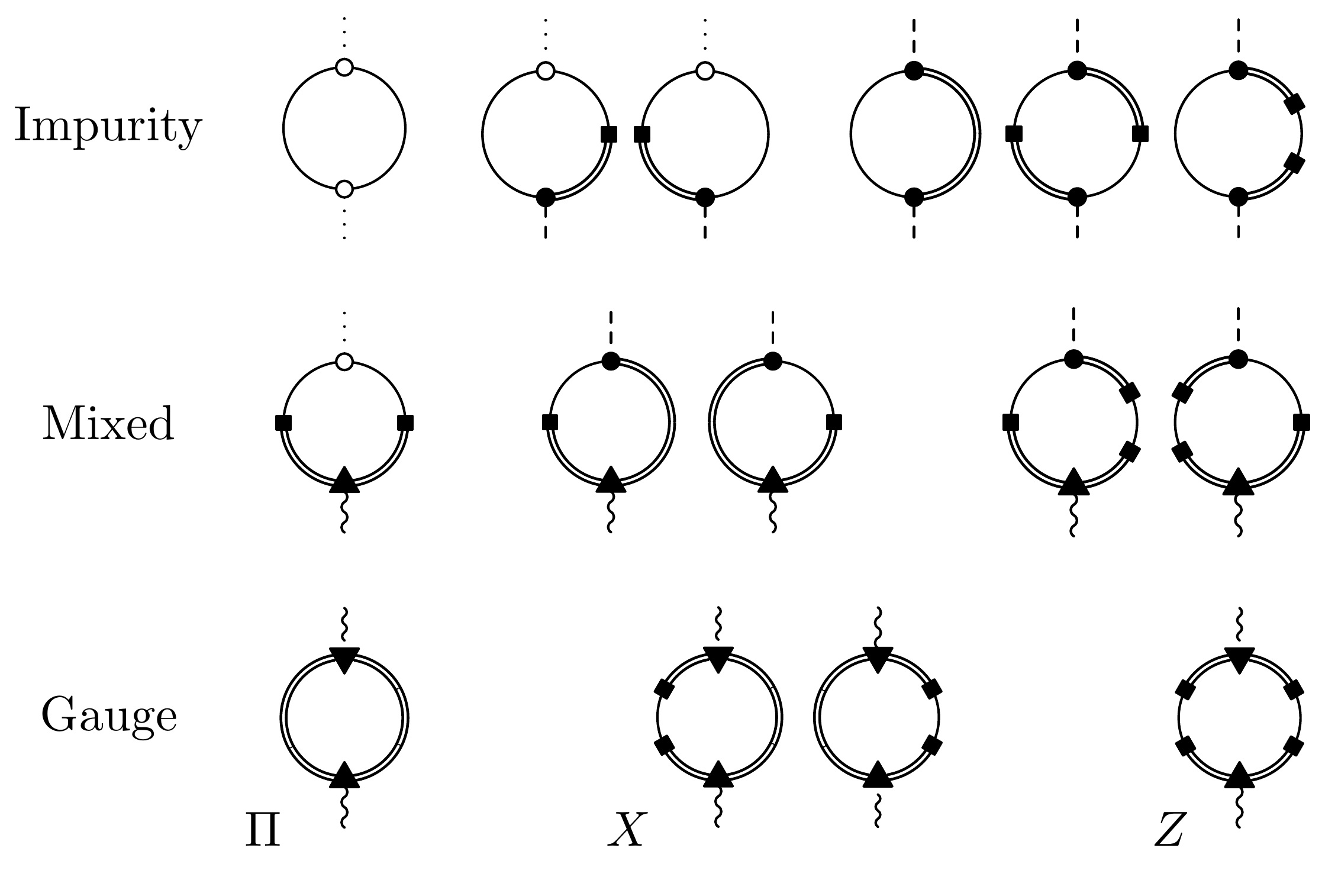}\caption{\label{fig:Diagrams}Bubble diagrams obtained summing over the fermionic
degrees of freedom. Due to parity mixed diagrams and are evaluated
to be zero as well as the one labeled by$Z$.}

\end{figure}

Fluctuations due to the bosonic fields are obtained summing the fermionic
bubbles in the RPA approximation. Without the Kondo term $\left(J_{K}=0\right)$
the propagator $D_{\mu\nu}=\Pi_{\mu\nu}^{-1}$ of the longitudinal
and transverse components of the Gauge field is given by the density-density
and current-current response functions. Using the Coulomb gauge $\boldsymbol{\nabla}.\boldsymbol{a}=0$
the longitudinal part is fully gaped $\Pi_{00}\simeq\frac{m}{2\pi}$
yielding to screening, by the spinons, of a $U(1)$ test-charge. Therefore
one can safely ignore the dynamics of $a_{0}$. The transverse component
$\Pi_{i,j}=\left(\delta_{i,j}-\frac{q_{i}q_{j}}{q^{2}}\right)\Pi$
is gap-less and results from the Landau damping of the collective
transverse modes by the gap-less spinons. For $\left|\Omega_{u}\right|<v_{F}q$
we can write \begin{eqnarray}
\Pi\left(i\Omega_{u},\mathbf{q}\right) & = & \gamma\frac{\left|\Omega_{u}\right|}{q}+\chi q^{2}\end{eqnarray}
where $\gamma=\frac{k_{F}}{\pi}$ and $\chi=\frac{1}{12\pi m}$ \cite{Nave_2007}.
\\
Using the diagrammatic rules of Fig. \ref{fig:Vertices} the bubble-like
diagrams, including transverse gauge as well as $\kappa$ and $\lambda$
fluctuations, are given in Fig. \ref{fig:Diagrams} and are divided
in pure impurity diagrams, mixed diagrams and gauge diagrams. The
transverse component of the gauge vertex is such that $\mathbf{q}\times\boldsymbol{j}=-\frac{1}{m}\mathbf{q}\times\mathbf{k}$.
\\
\\
The impurity diagrams corresponding to the fluctuations of $\kappa$
and $\lambda$ were obtained in ref. \cite{Read_1983} and are given
explicitly in the Appendix \ref{sec:Imp_fluct}. %
\begin{figure}[H]
\centering{}\includegraphics[width=0.7\columnwidth]{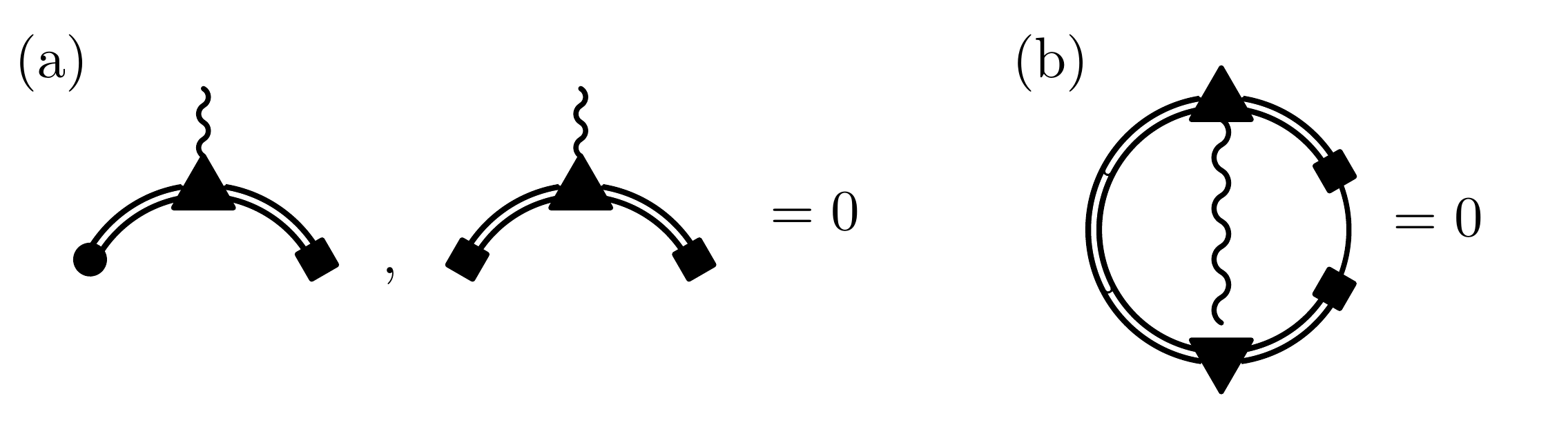}\caption{\label{fig:Other_Diagrams}(a) Pieces of diagrams giving zero by parity.
(b) $X$ Diagram contributing to the free energy. }

\end{figure}
It is easy to prove that due to parity considerations all the diagrams
including the pieces of Fig.\ref{fig:Other_Diagrams}-(a) are zero
(see Appendix \ref{sec:zero_diag}). These include the mixed diagrams
as well as the ones labeled by $Z$ in Fig. \ref{fig:Diagrams}. This
implies that at this order in the $1/N$ expansion the gauge propagator
decouples from the impurity degrees of freedom and the influence of
the impurity scattering enters only true the $X$ diagrams contribution:\begin{eqnarray}
 &  & X_{i,j}\left(i\Omega_{u},-\mathbf{q},\mathbf{q}'\right)=\frac{1}{\beta}\sum_{n}\int\frac{d\mathbf{k}}{\left(2\pi\right)^{2}}\times\\
 &  & b\kappa_{0}^{2}F_{0}\left(i\omega_{n}\right)G_{0}\left(i\omega_{n},\mathbf{k}-\mathbf{q}\right)G_{0}\left(i\omega_{n},\mathbf{k}-\mathbf{q}'\right)\times\nonumber \\
 &  & \left\{ \frac{1}{m}+\frac{\left(\mathbf{k}\times\mathbf{q}\right)_{i}\left(\mathbf{k}\times\mathbf{q}'\right)_{j}}{m^{2}\abs{\mathbf{q}}\abs{\mathbf{q}'}}\left[G_{0}\left(i\omega_{n-u},\mathbf{k}\right)+G_{0}\left(i\omega_{n+u},\mathbf{k}\right)\right]\right\} .\nonumber \end{eqnarray}
Taking into account all non-zero terms the bosonic action (\ref{eq:bosonic_action}),
developed at Gaussian order, writes now $s_{b}=s_{0}+\frac{1}{N}\left(s_{imp}+s_{a}\right)$,
where \begin{eqnarray}
s_{0} & = & -\tr\ln\left[-G_{0}^{-1}\right]-\tr\,\ln\left[-F_{0}^{-1}\right]\nonumber \\
 &  & +\frac{\beta}{J_{K}}\kappa_{0}^{2}-\varepsilon_{c}\frac{Q_{f}}{N}\beta+\beta\frac{V}{b}\mu\frac{Q_{f}}{N}\end{eqnarray}
is the value of the action at the saddle-point, $s_{imp}$ includes
the fluctuations of the impurity degrees of freedom (given the original
Read and Newns paper \cite{Read_1983} and in Appendix \ref{sec:Imp_fluct}
) and $s_{a}$ is the action for the transverse component of the gauge
field\begin{eqnarray*}
s_{a} & = & \frac{1}{\beta}\sum_{u}\int\frac{d\mathbf{q}\, d\mathbf{q}'}{\left(2\pi\right)^{4}}\,\bar{a}\left(i\Omega_{u},\mathbf{q}\right)a\left(i\Omega_{u},\mathbf{q}'\right)\times\\
 &  & \left[\delta\left(\mathbf{q}-\mathbf{q}'\right)\Pi\left(i\Omega_{u},\mathbf{q}\right)+X\left(i\Omega_{u},-\mathbf{q},\mathbf{q}'\right)\right],\end{eqnarray*}
where $\boldsymbol{a}=\hat{\mathbf{q}}_{\perp}a$ and $X_{i,j}=\left(\hat{\mathbf{q}}_{\perp}\right)_{i}\left(\hat{\mathbf{q}'}_{\perp}\right)_{j}X$.

\subsection{Specific Heat Capacity\label{sub:Specific-Heat-Capacity}}

We compute the specific heat considering the temperature dependency
of the free energy \begin{eqnarray}
F & = & -\frac{1}{\beta}\ln Z\nonumber \\
 & = & \frac{N}{\beta}\left\{ s_{0}+\frac{1}{N}\left(\frac{1}{2}\tr\ln\Gamma+\frac{1}{2}\tr\ln\left[\Pi+X\right]\right)\right\} .\label{eq:f}\end{eqnarray}
The $s_{0}$ term gives the contribution to the free energy of the
bulk fermionic spinons $\left(C_{v}^{\text{(\text{spinon})}}=N\frac{\pi^{2}}{3}Vn(0)T\right)$
and the leading order impurity term $\left(C_{v}^{(\text{imp})}=N\frac{\pi^{2}}{3}\rho(0)T\right)$.
The $1/N$ terms carry the free energy contributions from the bosonic
degrees of freedom. For low temperature all internal (fermionic) propagators
of Fig.\ref{fig:Diagrams} can be computed at $T=0$ and the temperature
dependency is given by the bosonic degrees of freedom \cite{Read_1983}.
The first next to leading order correction due to the impurity bosons
(proportional to $\tr\ln\Gamma$ in Eq. \ref{eq:f}) has been shown
to give a correction to the impurity contribution to the specific
heat \cite{Read_1983}. Defining $C_{v}^{(\text{imp})}=\gamma_{\text{imp}}T$
one obtains $\gamma_{\text{imp}}=\left(N-1\right)\frac{\pi^{2}}{3}\rho(0)$
which can be interpreted as the suppression of one of the $N$ impurity
degrees of freedom due to the existence of a constrain. \\
\\
Since the fluctuations of the gauge and impurity factorize new
phenomena can only arise from the $X$ corrections to the propagator
of the gauge. In a system with a dilute number of magnetic scatters
this term would be of the order of the density of impurities, in this
case of a single impurity it is simply proportional to $1/V$. It
is therefore natural to expand $\tr\ln\left[\Pi+X\right]=\tr\ln\left[\Pi\right]+\tr\left[\Pi^{-1}X\right]+...$.
The first term in the expansion is responsible for the gauge field
contribution to the specific heat $C_{v}^{(\text{gauge})}\propto V\left(\frac{\gamma T}{\chi}\right)^{2/3}$.
The correction to the free energy $\Delta F=\frac{1}{\beta}\frac{1}{2}\tr\left[\Pi^{-1}X\right]$
is given by the diagram of Fig. \ref{fig:Other_Diagrams}-(b). It
is easy to prove that such contribution vanishes remarking that one
can rewrite it as \begin{eqnarray}
 &  & \frac{1}{\beta}\tr\left[\Pi^{-1}X\right]=\frac{1}{\beta}\sum_{n}\int\frac{d\mathbf{k}}{\left(2\pi\right)^{2}}\times\\
 &  & 2b\kappa_{0}^{2}F_{0}\left(i\omega_{n}\right)G_{0}\left(i\omega_{n},\mathbf{k}\right)G_{0}\left(i\omega_{n},\mathbf{k}\right)\Sigma_{f}\left(i\omega_{n}\right)=0\nonumber \end{eqnarray}
where $\Sigma_{f}\left(i\omega_{n}\right)\propto-i\sign\left(\omega_{n}\right)\abs{\omega_{n}}^{\frac{2}{3}}$
is the spinon self-energy \cite{Mross_2010} given in Fig. \ref{fig:Propaga}-(b).
The vanishing of such term is a consequence of the independence of
$\Sigma_{f}$ from the spinon momentum. \\
\\
Thus the only correction to the specific heat due to the presence
of the impurity is given by a correction to $\gamma^{(\text{imp})}$
since all other terms vanish either by parity considerations of by
the above argument.

\subsection{Spin Susceptibility\label{sub:Spin-Susceptibility}}

In this section we consider the local spin-spin correlations at the
impurity site and its different contributions coming from the impurity-impurity
$\chi_{\text{imp},\text{imp}}\left(\tau\right)=\av{\mathbf{S}(\tau).\mathbf{S}(0)}$
, impurity-spinon $\chi_{\text{imp},\text{spinon}}\left(\tau\right)=\av{\mathbf{S}(\tau).\mathbf{S}_{f,(\mathbf{x}=0)}(0)}$
and from the local spinon-spinon $\chi_{\text{spinon},\text{spinon}}\left(\tau\right)=\av{\mathbf{S}_{f,(\mathbf{x}=0)}(\tau).\mathbf{S}_{f,(\mathbf{x}=0)}(0)}$
susceptibilities. In order to investigate the role of the impurity
and gauge degrees of freedom we consider the $1/N$ corrections of
the propagators and external vertices. \\
\begin{figure}[H]
\centering{}\includegraphics[width=1\columnwidth]{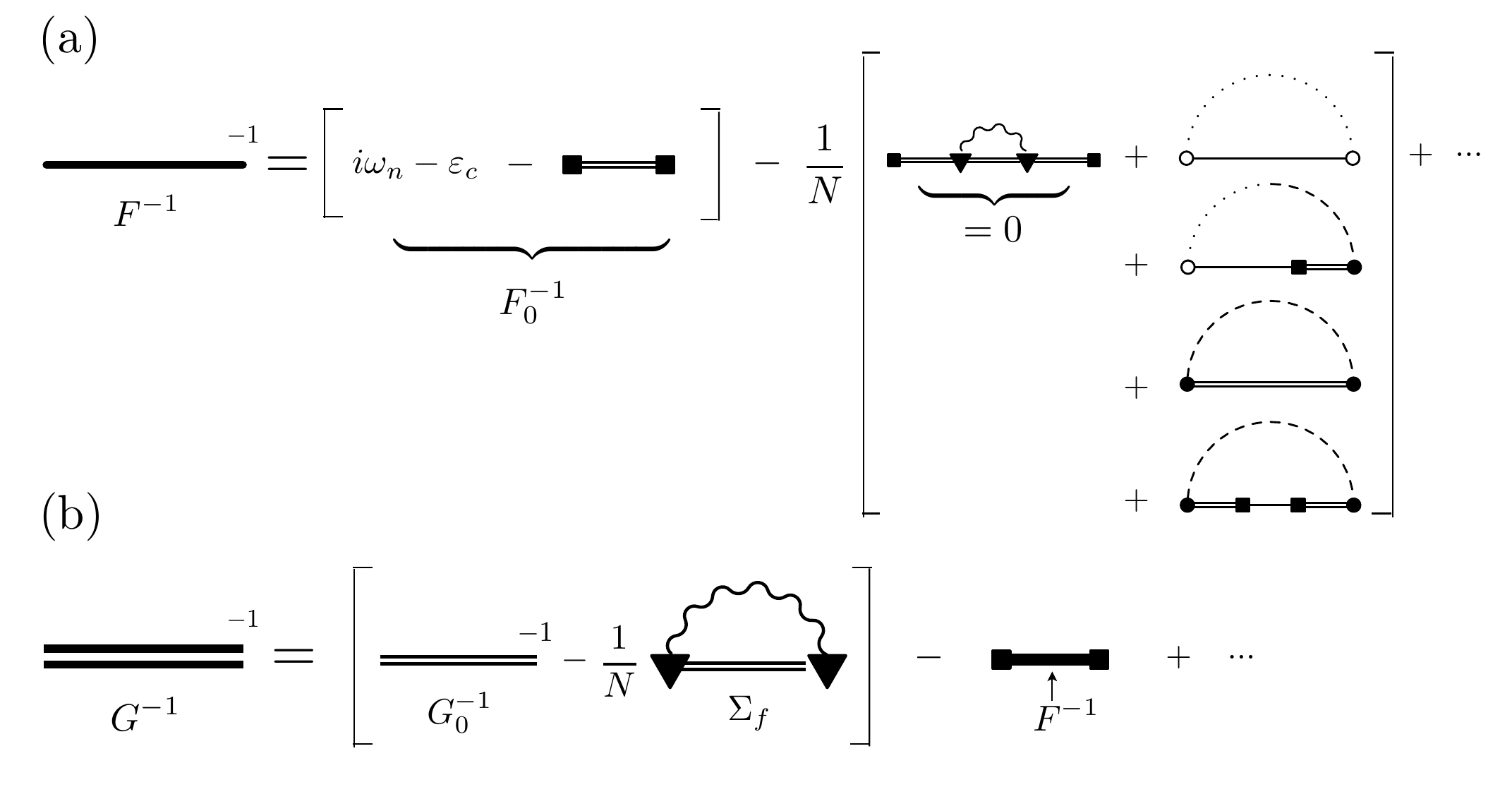}\caption{\label{fig:Propaga} Dyson's equation for the impurity (a) and spinon
(b) propagators up to order $1/N$, for sake of clarity symmetry related
diagrams are not shown. }

\end{figure}
Fig.\ref{fig:Propaga} shows diagrammatically the impurity and spinon
propagators up to order $1/N$. One can see that to this order the
impurity propagator has no corrections due to the presence of the
gauge field since terms like $\int d\mathbf{k}\ G_{0}G_{0}\Sigma_{f}=0$
vanish as a consequence of the independence of $\Sigma_{f}$ from
the spinon momentum. Alternatively one can use the renormalized spinon
propagator to compute the self energy of the impurity (second term
of $F_{0}^{-1}$ in Fig.\ref{fig:Propaga}-(a)) which would correspond
to a rearrangement of the terms in \ref{fig:Propaga}-(a) leading
to the same result. The impurity propagator is thus the same as if
the bulk was a regular Fermi-liquid. In this case one can use the
results in ref. \cite{Read_1983} where the fluctuations of the bosonic
impurity fields $\lambda$ and $\kappa$ were shown to renormalize
the Kondo temperature. \\
Besides the self energy term the spinon propagator, given in Fig.
\ref{fig:Propaga}-(b), has also a $1/V$ contributions from impurity
scattering, these can however be safely ignored in the computation
of the local susceptibility since it would give a $1/V^{2}$ correction.
\\
\begin{figure}[H]
\centering{}\includegraphics[width=1\columnwidth]{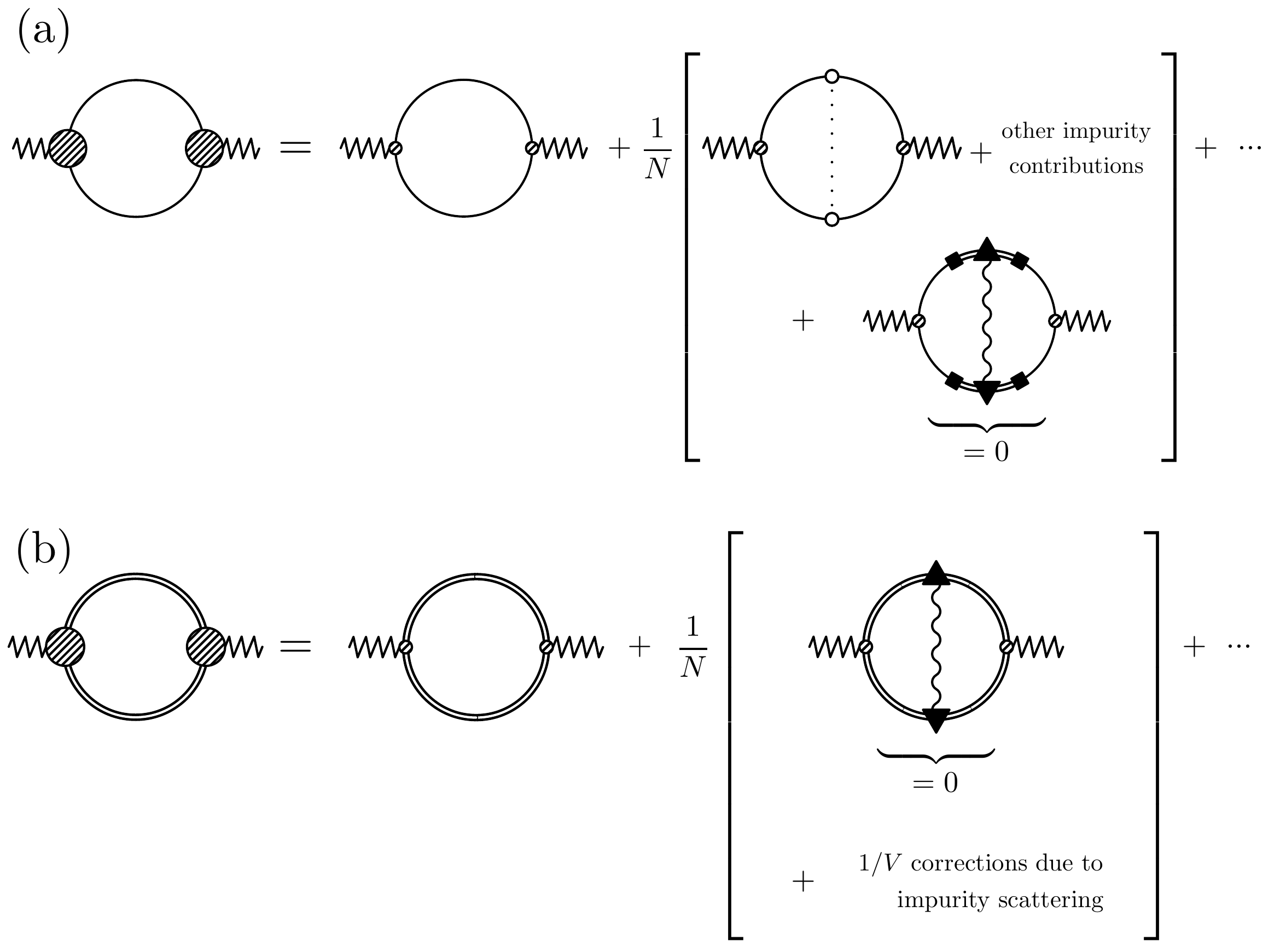}\caption{\label{fig:vertex-corr} Impurity-impurity and local spinon-spinon
spin susceptibilities. The vertex corrections are given to order $1/N$,
the bosonic propagators are obtained inverting the bubble like diagrams
of Fig. \ref{fig:Diagrams}. The external zigzag line carry spin and
frequency indices. }

\end{figure}
The impurity-impurity susceptibility is given at leading order in
$1/N$ by the bubble diagram of Fig.\ref{fig:vertex-corr}-(a) (first
term in the r.h.s.). $1/N$ vertex corrections due to the gauge field
arising in the impurity-impurity susceptibility also vanish (see Fig.\ref{fig:vertex-corr})
since they contain the terms like the ones in Fig.\ref{fig:Other_Diagrams}-(a).
One thus concludes that the impurity-impurity susceptibility $\chi_{\text{imp},\text{imp}}$
has no contribution from the gauge field at this order in $1/N$.
So the impurity degrees of freedom see only the local density of the
spinons, in particular the result given in \cite{Read_1983} for the
static susceptibility hold: $\chi_{\text{imp},\text{imp}}\left(i\Omega_{n}=0\right)=\frac{1}{3}J(J+1)(2J+1)\rho\left(0\right)$
where $N=2J+1$.\\
\\
Gauge contributions are known to enhance Friedel-like oscillations
in $U(1)$ spin-liquids \cite{Altshuler_1994,Kolezhuk_2006,Mross_2010},
this is a consequence of the renormalization of the $2k_{F}$ component
of the susceptibility vertex. One could thus expect that the local
spinon-spinon susceptibility carried some trace of this behavior.
Remarkably no vertex corrections to the local susceptibility due to
the gauge field are possible since simple parity arguments like the
one used in Appendix \ref{sec:zero_diag} show that the contribution
given by the second diagram in the r.h.s of Fig.\ref{fig:Other_Diagrams}-(b)
vanishes. \\
\\
Finally the crossed impurity-spinon susceptibility can also be
shown to remain unaffected by the presence of the gauge field using
the same simple arguments. \\
This shows that the local measurements of the susceptibility at
the impurity site are insensitive to the gauge degrees of freedom.

\section{Discussion\label{sec:Discussion}}

We considered the Kondo screening in a bulk system of spinons strongly
interacting with a $U(1)$ gauge field. While it is remarkable that
Kondo screening can occur for a charge insulator, the results obtained
here predict that no particular signature due to the presence of the
gauge field can be measured if only the impurity degrees of freedom
or local magnetic properties are monitored at the impurity site. \\
\\
The presence of the impurity destabilizes spin-liquid phase locally
and Friedel-like oscillations are expected once the density of spinons
is locally disturbed. This is due to the last term in \ref{fig:Propaga}-(b),
however they would equally be present if the density of spinons was
changed by non-magnetic means as for example at the sample edges or
near non-magnetic impurities. Such oscillations should be enhanced
by the presence of the gauge fields \cite{Mross_2010} however they
are non-local measures. Local probes will be incapable of distinguish
the bulk system from a Fermi-liquid. In particular the Wilson ratio
$R=\frac{\pi^{2}\chi_{\text{imp},\text{imp}}\left(0\right)}{J(J+1)\gamma_{imp}}=\frac{N}{N-1}$
for this case is the same as for a magnetic impurity embedded in a
Fermi-Liquid \cite{Read_1983}. \\

\begin{acknowledgments}
We thank T. Senthil for helpful discussions. PR acknowledges support
through FCT BPD grant SFRH/BPD/43400/2008. PAL acknowledges the support
by DOE under grant DE-FG02-03ER46076. 
\end{acknowledgments}
\appendix

\section{Some details }

\subsection{Prove that the diagram of Fig.\ref{fig:Other_Diagrams}-(a) is zero\label{sec:zero_diag}}

\begin{figure}[H]
\centering{}\includegraphics[width=0.3\columnwidth]{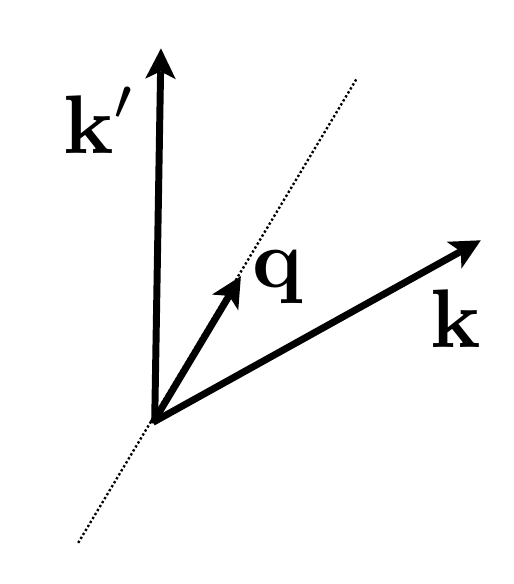}\caption{\label{fig:reflect_k} Change of variables that implies that the diagram
in Fig.\ref{fig:Other_Diagrams}-(a) is zero. }

\end{figure}

The diagram of Fig. \ref{fig:Other_Diagrams}-(a) is given by \begin{eqnarray*}
W & = & b\kappa_{0}^{2}\int\frac{d\mathbf{k}}{\left(2\pi\right)^{2}}G_{0}\left(i\omega_{n},\mathbf{k}\right)\frac{\left(\mathbf{k}\times\mathbf{q}\right)}{m\abs{\mathbf{q}}}G_{0}\left(i\omega_{n-u},\mathbf{k}-\mathbf{q}\right).\end{eqnarray*}
Changing variables $\mathbf{k}\to\mathbf{k}'=2\frac{\mathbf{q}.\mathbf{k}}{\mathbf{q.q}}\mathbf{q}-\mathbf{k}$,
where $\mathbf{k}'$ is obtained reflecting $\mathbf{k}$ on axes
$\mathbf{q}$ (see Fig.\ref{fig:reflect_k}), leaves the norms $\abs{\mathbf{k}}=\abs{\mathbf{k}'}$
and $\abs{\mathbf{k}-\mathbf{q}}=\abs{\mathbf{k}'-\mathbf{q}}$ invariant
and changes the sign of $\mathbf{k}\times\mathbf{q}=-\mathbf{k}'\times\mathbf{q}$.
Since $G_{0}\left(i\omega_{n},\mathbf{k}\right)=G_{0}\left(i\omega_{n},\abs{\mathbf{k}}\right)$
for a spherically symmetric Fermi surface it follows that $W=-W=0$.

\subsection{Impurity Fluctuation \label{sec:Imp_fluct}}

The impurity action at Gaussian level is given by

\begin{eqnarray*}
s_{imp} & = & \frac{1}{\beta}\sum_{n}\frac{1}{2}\left[\begin{array}{c}
\bar{\kappa}(i\Omega_{u})\\
\bar{\lambda}(i\Omega_{u})\end{array}\right]^{T}\boldsymbol{\Gamma}(i\Omega_{u})\left[\begin{array}{c}
\kappa(i\Omega_{u})\\
\lambda(i\Omega_{u})\end{array}\right]\end{eqnarray*}
where \begin{eqnarray*}
\boldsymbol{\Gamma}(i\Omega_{u}) & = & \left[\begin{array}{cc}
\frac{\delta^{2}s}{\delta\bar{\kappa}\delta\kappa} & \frac{\delta^{2}s'}{\delta\bar{\lambda}\delta\kappa}\\
\frac{\delta^{2}s'}{\delta\bar{\lambda}\delta\kappa} & \frac{\delta^{2}s'}{\delta\bar{\lambda}\delta\lambda}\end{array}\right]\end{eqnarray*}
is the fluctuation matrix. If one evaluate the fermionic Matsubara
sums at zero temperature its entries are given by

\begin{eqnarray*}
\delta_{\bar{\lambda}}\delta_{\lambda}s & = & \frac{\Delta}{\pi\left|\Omega_{n}\right|\left(2\Delta+\left|\Omega_{n}\right|\right)}\ln\left[\frac{\varepsilon_{c}^{2}+\left(\left|\Omega_{n}\right|+\Delta\right)^{2}}{\varepsilon_{c}^{2}+\Delta^{2}}\right];\\
\delta_{\bar{\lambda}}\delta_{\kappa}s & = & \frac{2ibn(0)\kappa_{0}}{\left|\Omega_{n}\right|}\left[\tan^{-1}\left(\frac{\left|\Omega_{n}\right|+\Delta}{\varepsilon_{c}}\right)-\tan^{-1}\left(\frac{\Delta}{\varepsilon_{c}}\right)\right];\\
\delta_{\bar{\kappa}}\delta_{\kappa}s & = & bn(0)\left(\frac{2\Delta}{\left|\Omega_{n}\right|}+1\right)\ln\left[\frac{\varepsilon_{c}^{2}+\left(\left|\Omega_{n}\right|+\Delta\right)^{2}}{\varepsilon_{c}^{2}+\Delta^{2}}\right].\end{eqnarray*}

\end{document}